\documentclass[prd,nobibnotes,nofootinbib,showpacs,12pt]{revtex4}
\usepackage[papersize={8.5in,11in}]{geometry}
\usepackage{datetime}
\usepackage{amsmath} 
\usepackage{amssymb}
\usepackage{amsfonts}   
 \DeclareMathOperator{\Tr}{Tr}
\usepackage{graphicx}   
\usepackage{verbatim}   
\usepackage{color}      
\usepackage{subfigure}  
\usepackage{hyperref}   
\usepackage[english,spanish,romanian]{babel}
\usepackage{combelow}
\begin{document}

\title{Lattice radial quantization by cubature.}
\author{Herbert Neuberger}
\email{neuberg@physics.rutgers.edu}
\affiliation{Department of Physics and Astronomy, Rutgers University,\\ 
Piscataway, NJ 08855, U.S.A} 

\begin{abstract}
Basic aspects of a program to put 
field theories quantized in radial coordinates
on the lattice are presented. Only scalar fields
are discussed. Simple examples are solved to illustrate 
the strategy when applied to the 3D Ising model. 
\end{abstract}

\date{November, 12, 2014}

\pacs{11.15.Ha, 11.25.Hf}

\maketitle

\section{Introduction}

This paper is a continuation of a feasibility 
study~\cite{ourplb} carried out with 
Brower and Fleming where we put the path integral corresponding
to the radially quantized version of a putative conformal
field theory (CFT) 
on the lattice and then calculated numerically some
eigenvalues of the transfer matrix, ${\cal T}$, in the 
$t=\log r$ direction. Here $r$ is the flat Euclidean distance from
a selected point, the origin. The units of $r$ are irrelevant. 
The spectrum of $\log {\cal T}$ contains the
scaling dimensions of the scaling fields. 
We were mainly motivated by work by Cardy~\cite{cardy}. 
The main observation of~\cite{ourplb} was that 
spectral regularities characteristic of a CFT could
be used to determine non-universal scale factors. 

Some couplings 
need to be adjusted in order that the IR regime fall 
into a desired universality class. In radial 
quantization one needs to employ a variation on classical 
flat space methods to tune into criticality. 
The numerical application of~\cite{ourplb} was to a piecewise flat 
deformation of the sphere and 
classical flat space methods could be used.
Radial quantization sacrifices $d$-component translations in exchange to 
preserving dilatation at the UV-cutoff level. 
The role of the flat space mass term is taken by
a ``mass'' term that now explicitly breaks translational invariance but
preserves dilatations.

Infinite towers of equally spaced levels 
permeating the spectrum are the most prominent 
spectral regularity of a CFT. The spacing is the 
same for all towers. I restrict my
attention to CFT's with an energy-momentum tensor.
Spectrally, this means that there exists a state
transforming as a traceless symmetric second rank
tensor whose dimension is known in advance to be 
$d$ times the universal level spacing in the towers. 
The lowest level in a tower is called ``primary'' 
and the higher members are its ``descendants''. 
The subspace spanned by the tower is invariant 
under the conformal group. Larger irreducible
multiplets of $O(d)$, where $d$ is the dimension of 
spacetime, appear at higher levels in the tower, sharing
average spectral weight between distinct towers. Exponentially
growing degeneracies appear asymptotically. 

Any discretization of 
the sphere will break continuum
$O(d)$ to a finite group $Q'$. I am only considering 
non-abelian $Q'$'s and focus on the largest ones.  
Each element of $Q'$ 
can be written as the product of an element of a $Z_2$ and
an element of $Q$ where $Q$ is a nonabelian 
subgroup of $SO(d)$ and the non-trivial element of 
$Z_2$ takes a point on the sphere to its diametrical image. 
I assume $d\ge 3$. Then, the number of elements in $Q$, $|Q|$, 
is bounded. In $d=3$ we shall work with $Q=I$, the largest finite
nonabelian discrete subgroup of $SO(3)$. $Q$ has 60 elements and
$Q'=Q\times Z_2$ is the 120 element group $I_h$. $I$ is isomorphic 
to $A_5$ the group of even permutations of 5 objects. $A_5$ is
a subgroup of $S_5$, the group of all permutations of 5 objects and
$S_5=Z_2\ltimes A_5$. So, $I_h$ is not isomorphic to $S_5$. 
$I$ and $I_h$ are Icosahedral groups. The double Icosahedral 
group is not needed for scalar fields. 

The breaking of $SO(d)$ splits multiplets of higher multiplicity but
maintains a few small ones. A few low rungs
of ladders making up the towers 
corresponding to low scaling dimension primaries
can be identified. Small dimensions are at the bottom 
of the entire spectrum where average degeneracy is low, facilitating 
identification. 
One can imagine a sequence of
adjustments on the action that zero out the splits of
larger $SO(d)$ multiplets one by one as one ascends in level. 
I shall later show a way to
do this. Eliminating these splits restores
continuum rotational invariance at the spectral level. 
This is not ``fine tuning'', 
but rather ``improvement''. (The restoration 
of rotational invariance in a stochastic way has
been a topic in lattice field theory in 
the eighties~\cite{randlat}; this is an
option I ignore in this work.) 
Eliminating the
splits does not produce equal spacings. 
Fine tuning to criticality 
is the adjustment needed to get a few low lying 
spacings equal 
to each other and similarly 
correct dimensions for the energy-momentum tensor. 
The equal spacing property should spread 
upwards into the spectrum 
as the lattice is refined. 
It is not clear in advance how much tuning is required
to achieve this. 

I would like to define a transformation and a space
of Hamiltonians acting on a common Hilbert space 
which would produce the right 
value for the energy-momentum state 
dimension in units of tower spacings
upon infinite iteration if the initial Hamiltonian
is tuned. 
This transformation would be an analogue of the RG
transformation in flat space. In flat space  
dilatation invariance gets restored 
at the fixed points. In radial quantization 
translations do; this is why tower spacings and 
the dimension of the 
energy-momentum tensor get tied together. It 
may turn out that there are some differences between
this analogue and the standard RG. This might 
be of interest in Particle Physics by expanding 
the concept of ``fine tuning''.  

In practice, my program is anchored on the existence of 
an action and the ``elementary'' fields which make it 
up. The concept of ``elementary
field'', to say nothing about ``action'', is not fundamental. 
But, so long as one works
within a framework that provides in principle a 
constructive approach to the final continuum quantum 
system one needs to start at some corner which is under
control, albeit devoid of fundamental significance. 
My corner is a well defined  
replacement of the formal path
integral. It has an integration measure, and 
the integrals of a wide class of functionals of the elementary field exist.

Much of the subsequent discussion and all the examples are 
in $d=3$. 
The way rotational invariance is violated differs from 
flat space, where translations play a  
fundamental role and the spacetime group is 
a semidirect product of translations and $Q$. 
Lattice translational 
invariance guarantees that the specific
$Q$ associated with the grid choice acts as a symmetry
with respect to any vertex chosen as origin. 
Local lattice densities fall into multiplets under $Q$. 
In radial quantization the origin is fixed once and for 
all. Densities localized
at few selected points on the sphere might fall into
multiplets of some subgroup of $Q$. 
The number of points this can happen at is a 
divisor of $|Q|$. The eigenstates of the transfer matrix ${\cal T}$ 
do fall into $Q$-multiplets but correspond to global states w.r.t. the sphere. 
The CFT 
state-operator correspondence depends crucially on 
translational invariance. 

Understanding the difficulty with rotational invariance, 
we identified two options to choose from.
The first, adopted in~\cite{ourplb}, is to 
replace the continuum field theory on the sphere by
the same continuum field theory  on another space. 
Intuition and evidence 
from previous numerical work, indicated that 
low lying spectral properties 
of the deformed theory match closely those expected in  
the radial case. In some sense the two theories
seem connectible in a way expressible by a converging perturbative expansion. 
In the deformed theory the sphere $S_{d-1}$
gets replaced by an almost everywhere flat manifold, 
with flat simplicial patches glued to each other
to make up a space of spherical topology. For clarity, 
I now set $d=3$. Imagining a paper model of some
polyhedron with triangular faces one recognizes 
singularities at the vertices, points where more than
2 triangles meet. The induced metric is now 
flat everywhere except at the vertices. There
are no singularities away from the vertices 
even on the edges, because the paper can be 
flattened out at the fold. The singularity is a 
cone singularity; one can cut out a vertex and glue
back smoothly a paper cone in its stead. 
The cones have 
angle deficits that add up to the area of the sphere. 
The group $Q$ acts transitively on the cones. 
 
It was natural to explore this option first 
even if we ultimately insisted 
to work on the sphere.  The most natural choice on the
sphere is to model the action on the
finite elements method (FEM)~\cite{fix}. In FEM one is 
working on a space as above, only the number of
cones increases with the number
of vertices. One can ensure that $Q$ still
permutes the cones, but the action would not 
be transitive. One does not really 
escape conical singularities in FEM. One needs
to show that their effect becomes sub-leading as the
number of vertices increases in a chosen specific 
prescription. This is plausible and progress in that direction 
has been described in lattice conference
contributions in 2013~\cite{brower2013} and 2014~\cite{brower2014}. It seems to me unlikely 
that sub-leading corrections would organize themselves 
by scaling dimensions of irrelevant scaling fields like on flat lattices. 
Piecewise flat spaces do approximate smooth manifolds 
in a well defined manner~\cite{cheeger}, but the 
issue is subtle~\cite{wardetzky}. Subtleties
were identified a long time ago~\cite{sorkin}. 
Similar problems, in
particular for the case of the two sphere, appear
for example in applications to climate control, 
medical imaging and fluid dynamics~\cite{shallow},~\cite{imaging}.

The continuum 
formulation of~\cite{ourplb} has the advantage that
keeping the number of conical singularities fixed 
preserves as much rotational symmetry as possible and
simultaneously preserves {\it infinitesimal} 
translational symmetry away from the 
singularities. In turn, this gives
an energy-momentum tensor whose divergence is zero 
except on the cone lines (traced out in $t$ by 
the vertices), where 
singular sources reside. It is plausible that for
appropriate bulk quantities the contribution of the 
cone singularities are sub-dominant in the IR. 
Having large swaths of flat space makes it possible 
to use well tried methods to tune into criticality.
This separates the problem of using expected spectral 
regularities for establishing criticality from 
exploiting them for numerical determination 
of critical exponents when criticality is assured 
independently. 
Criticality was determined in~\cite{ourplb} by
numerically studying how the probability 
distribution of the
order parameter behaved as the number of vertices increased
with the help of Binder's cumulant~\cite{binder}.
The results verified that indeed the cones made no
contribution because the powers involved matched 
well against known values from conventional 
flat space studies~\cite{vicari}. These findings confirmed earlier work
with cubic symmetry~\cite{weigel}. No
attempt was made to tune the strength of the singularities.
I do not know whether it would have been possible to
tune the couplings attached to the lattice vertices 
at the cone singularities 
to values that would have zeroed out the lowest 
$SO(d)$ multiplet split we found in the odd sector
of the model at $l=3$~\cite{ourplb}. 
Studies of cone singularities in other contexts indicate
that one parameter should suffice because this is the
freedom one has when extending the Laplacian action 
to the singularities~\cite{cones}. 
Symmetrical arrangements of cones also appear 
in classical general relativity in the context of
symmetrical arrangements of cosmic strings~\cite{cosmic}.

The obvious disadvantage of the option chosen in~\cite{ourplb}
is that the connection to the spherical case needs to
be fully understood. I think this will happen. 
How well this would work quantitatively 
is premature to speculate. The numerical 
indication from~\cite{ourplb} is that it 
should work well in the 3$d$ Ising model. 

I believe that learning how to deal non-perturbatively 
with field theory on
classical curved backgrounds is a promising research 
direction for non-QCD oriented 
lattice field theory. Lattice radial quantization is one example.
The present paper is both elementary and detailed. The intent is to
make it easily accessible specifically to lattice theorists 
among other readers. 
Subsequent results from this program will
hopefully be less elementary and more succinctly presented. 

In the next section I shall describe the application of the
cubature framework to constructing lattices and actions. 
Cubature is the higher dimension generalization of Gaussian quadrature.
The cubature framework is introduced as an 
alternative to FEM, the natural first choice. I shall get back to 
compare these two viewpoints later in the paper. 
I have no information 
enabling me to compare the {\it effectiveness} of these two viewpoints. 

The cubature section is followed by a section in which the
transfer matrix is constructed. This makes it clear that one
has reflection positivity and also prepares the ground
for working out a toy example exactly, that is, without
any stochastic element in the method of solution. 

Rotation symmetry is then taken up in quite great detail in
the section that follows. Simple examples comprise the
last proper section, coming before the summary.

\section{Lattice action by cubature.}

In this work I shall look for an alternative to
FEM while working directly on the sphere.
Since any discretization is comparable to any
other this distinction may turn out to be just 
semantics.
Be that as it may, I think this alternate 
way of thinking will provide a procedure 
which differs substantially in details from the FEM route. 

Special
properties of the spectrum in the continuum 
theory will tell us how to tune the system so that 
its IR behavior falls into the desired universality class.
Relatively to~\cite{ourplb} I add the requirement that the energy momentum
tensor state be identified and that its energy be compatible with the spacing between the
primary and descendants for various primaries.  

I shall look at the discretization problem
from the viewpoint of cubature on the round sphere. 
The basic problem of cubature on the sphere~\cite{cub} deals with is constructing good 
approximations of the form
\begin{equation}
\int f({\hat\omega}) d\omega \approx \sum_{i=1}^N w_i f({\hat\omega}_i).
\label{cubature}
\end{equation}
$\hat\omega$ is a point of $S_{d-1}$ represented as unit vector in $R^d$, where $d$ is the spacetime dimension.
$d\omega$ is the measure on the sphere, normalized in
the standard manner. The $w_i$'s are weights, preferably
all positive. The points ${\hat\omega}_i$ reside on
the sphere. For a given $N$ we require the above
approximation to be exact for eigenfunctions of the
spherical Laplacian $-\partial^2_\omega$ 
from the lowest
level to a maximal level $\lambda_N$. As $N$ increases, $\lambda_N$ increases. The ${\hat\omega}_i$'s
are required to fall into complete orbits under 
the action of $Q$. In the
simplest case, one views the ${\hat\omega}_i$'s as fixed
and solves a linear equation for the $w_i$'s, looking
for the largest $\lambda_N$ that can be achieved. 
Alternatively, one may consider also the ${\hat\omega}_i$'s
as variables (subjected to $Q$-symmetry) and then
one has to solve a nonlinear system~\cite{gaussquad}. This extra work
is compensated by a larger $\lambda_N$ attainable 
at fixed $N$.

In our application $f({\hat\omega})$ is 
the continuum action density at a fixed $t$. The
$t$ direction is discretized in equal intervals in the
standard way. Since this discretizes $\log r$, the
expectation in~\cite{ourplb} was that this approach treats
the degrees of freedom in a way commensurate with their
contribution to the path integral at criticality. Therefore, 
this regularization would eventually turn out to be more
effective than the flat space one. In the following the ${\hat\omega}_i$'s are chosen first and the weights are 
found from linear equations. The continuum limit 
is expected to emerge as $N\to\infty$. 
I leave issues of efficiency of the implementation for
the future, after enough testing is carried out to
gain trust in the strategy.

The composite fields, including the action density, 
will be constructed out 
of one elementary scalar field $\phi$ which is defined 
at the points $(t_n,{\hat\omega}_i)$, 
$t_n=n\delta ,~n \in \mathbb{Z}$. Where possible, 
I shall suppress the $t$-coordinate
for simplicity. 
Reintroducing the $t$ dependence is a trivial matter. 
Accordingly, the variables of integration in 
the path integral are the $\phi({\hat\omega}_i)$.
They are thought of as coming from a function $\phi({\hat\omega})$, where $\hat\omega$ 
is continuous. The action density is a non-linear functional 
of $\phi$. The weights are fixed to reproduce exactly
the integrals of the action density when it is
limited to a finite number of low $l$ spherical waves.
One needs to work out what this means 
in terms of the spherical wave content of $\phi$.  

The continuum action is written in the form
\begin{equation}
A=\int d\omega d\omega^\prime \phi({\hat\omega})
K({\hat\omega},{\hat\omega^\prime}) \phi({\hat\omega^\prime}) +\int d\omega V(\phi({\hat\omega}))
\end{equation}
The kernel $K({\hat\omega},{\hat\omega^\prime})$ is 
the matrix element of an operator $K=f(\partial^2_\omega)$.
For scalar field theory a smooth 
UV cutoff can be introduced in the continuum 
action directly by implementing smearing~\cite{smear}.(For gauge theories, smearing requires 
a non-linear PDE, and is therefore introduced
only at the level of observables.)
\begin{equation}
K=(1-e^{s\partial^2_\omega})/s
\label{hk}
\end{equation}
The appearance of $e^{s\partial^2_\omega}$ is familiar to field theorists from
rigorous studies of the RG~\cite{gallavotti}. 
$K\to -\partial^2_\omega$ as $s\to 0$. 
$K$ is very popular in quite disparate fields of
science~\cite{belkin}. 
The UV cutoff is $1/\sqrt{s}$. After discretization
yet another UV cutoff enters, given by $\sqrt{N}$. One should choose $s=c/N$ with a constant $c$
of order of the area of $S_{d-1}$; convergence
can be tuned by adjusting this constant. 
$K$ is chosen as essentially an exponential because this 
produces a simple expression for
the $K({\hat\omega},{\hat\omega^\prime})$ in
any dimension. Explicitly, for $d=3$, 
\begin{equation}
<{\hat\omega}|e^{s\partial^2_\omega}|{\hat\omega^\prime}>=
\sum_{l=1}^\infty \frac{l(l+1)}{4\pi} e^{-sl(l+1)} P_l(
\cos({\hat\omega}\cdot{\hat\omega^\prime})),
\label{hkm}
\end{equation}
where $P_l$ is a Legendre polynomial in standard normalization. 	
The extra factor of $s$ in eq.~(\ref{hk}) is irrelevant
since we shall introduce an overall 
coupling $\beta_\omega >0$ in the integrand, $\exp(-\beta_\omega A)$, of the path integral for the partition
function. 

The discretized version of $A$ is
\begin{equation}
A=\sum_{i,j} w_i \phi({\hat\omega}_i)K({\hat\omega}_i,
{\hat\omega}_j) w_j \phi({\hat\omega}_j)+\sum_i w_i 
V(\phi({\hat\omega}_i)).
\end{equation} 
The weights $w_i$ depend only on the vertices $\hat\omega_i$ and not on the form of the action. 
The discretized action can be rewritten in shorter
form:
\begin{equation}
A=\sum_{i,j} w_i \phi_i K_{i,j} w_j \phi_j+\sum_i w_i 
V(\phi_i).
\end{equation}
The role of the unit operator in eq.~(\ref{hk}) is
to ensure that the kernel has zero as its lowest energy,
with constant eigenfunction. All other eigenvalues
are positive. This will hold for any symmetric 
matrix $K^\circ$ with positive off-diagonal terms and
diagonal terms determined by them.
\begin{equation}
K^\circ_{i,j} = \begin{cases} w_i K_{i,j} w_j &\mbox{if } i\ne j \\
-\sum_{k\ne i} w_i K_{i,k} w_k &\mbox{if } i=j. \end{cases}
\end{equation}
Thus, only matrix elements of the heat kernel between
unequal positions enter the discrete action. These terms
are all finite and have simple 
approximate expressions for $s\to 0$~\cite{mina}.

The discrete version of the action is no longer exactly 
equal to its continuum version even if the decomposition of $\phi$ 
contains only spherical wave functions with $l$ smaller than
some constant. We can arrange for the discrete and
continuum version to be numerically close to each other for 
a range of angular
momenta in the decomposition of $\phi$, $l\le\zeta\lambda_N$, 
($0<\zeta <1$ does not depend on $N$) if the non-linearity of the potential term is
polynomial.

First consider the quadratic term in the continuous action. It is obvious that
\begin{equation}
K({\hat\omega_1},{\hat\omega_2})= 
\sum_{l=0}^\infty\sum_{m=-l}^l Y_{lm}({\hat\omega_1}Y_{lm}^*({\hat\omega_2})\frac{1-e^{-sl(l+1)}}{s},
\end{equation}
where the $Y_{lm}$ are standard spherical harmonics.
I already explained that in eq.~(\ref{hk}) the discretization takes 
care of the 1 exactly and that the $1/s$-factor is
irrelevant. 
Hence the quadratic piece of the action can be taken as
\begin{equation}
{\cal Q}[\phi]=-\sum_{l,m} |{\cal L}_{lm}[\phi]|^2
e^{-sl(l+1)}~~{\rm with}~~
{\cal L}_{lm}[\phi]=\int Y_{lm}({\hat\omega})\phi({\hat\omega})d\omega .
\end{equation}
If $\phi$ decomposes into a sum of $l\le \lambda_N$ spherical waves, the linear functional 
${\cal L}_{lm}[\phi]$ will be exactly given by its discrete
counterpart for $l\le\lambda_N/2$ by
the rules of addition of angular momenta. 
The contribution to
the sum giving the quadratic functional ${\cal Q}[\phi]$
from terms with $l > \lambda_N/2$ will be relatively 
suppressed by $\sim e^{-s(l(l+1))}$. 
Choosing a sizeable value for $c$ in $s= c/N$ 
we can arrange for this correction to be small. 
Now consider the potential term in the action.
As long as $V$ is a polynomial of finite degree,
the continuum would agree with a discretized version
exactly with a $\zeta$ as above given by $\zeta=1/{\deg}(V)$. 

Note that there is no sharp cutoff in angular
momentum; such a sharp cutoff, while acceptable in principle, will have qualitative 
non-universal impact on the form of subleading 
power corrections in the IR. 
I do not know what the structure of these
would be for radial quantization. 
However, it is well known that
the effective Lagrangian treatment of 
subleading corrections in the 
approach to continuum in flat space 
breaks down if one uses a sharp momentum cutoff.

In the above construction, unlike
in the FEM case, the gradients of $\phi$ are not 
individually discretized; only the Laplacian is. 
In the case of FEM, the main step was to go from the
Laplace equation to a minimization problem which
required expressing the Laplacian as a sum of 
squares involving only 
first order derivatives obtained after 
an integration by parts in the action 
(the domain is finite).
The search for the minimum is carried out 
in the larger space of functions that have
piecewise continuous first order derivatives.
The jumps in the first order derivatives 
are integrable. (In the form of the action employing
the Laplacian, this would require dealing with 
$\delta$-function singularities in the integrand that
need to be discretized.)
The domain is decomposed into
flat pieces which become smaller and smaller and
one can prove that the solution to the
minimization problem converges to the
regular solution of the second order
PDE one started from~\cite{dziuk}. 
In the path integral, it is
not that important to have convergence of solutions
of the discrete variational problem to solutions of
the continuum PDE. We want the correlation functions to
converge, but the $\phi$ integration 
variables themselves are typically
quite rough. 

Discretizing the entire Laplacian at once, rather than
decomposing it as in continuum and discretizing
the individual terms clearly is a more general
approach as the discretized kernel no longer is
constrained to
admit any decomposition. An analogue 
strategy provides the single way known to date 
to discretize the exactly 
massless Dirac equation in the background of an arbitrary
lattice gauge field~\cite{overlap}.

 This concludes the general description of how the 
 path integral is defined by discretization.

\section{Transfer matrix.}

In the continuum the action is symmetric under $O(d)$ and $D$ which
consist of proper and improper rotations and dilatations. Up to a shift by 
the vacuum energy and an overall scale the scaling dimensions 
under $D$ are the spectrum of the transfer matrix ${\cal T}$. Another 
important symmetry is $\mathbb{I}^\circ$, inversion. It reverses the
sign of $t$. (The word inversion is
also used for the $Z_2$ generator extending $Q$ to $Q'$. 
Which is meant will be clear from the
context.)
This gives reflection positivity in the Euclidean
formulation and, therefore, unitary time evolution in  
Minkowski space~\cite{jaffe}.
 
The lattice action preserves $Q'$,
an infinite discrete subgroup of $D$, and $\mathbb{I}^\circ$. The lattice
action is
\begin{equation}
\begin{aligned}
A[\phi]=&\delta\sum_{i=1}^N\sum_{j=1}^N \sum_{n=-\infty}^\infty w_iw_j \left [ \phi_{n,i} K_{i,j} \phi_{n,j}\right ]+
\\&\frac{\beta_t}{2\delta\beta_\omega}\sum_{i=1}^N \sum_{n=-\infty}^\infty w_i (\phi_{n,i}-\phi_{n-1,i})^2+\frac{\kappa\delta}{2\beta_\omega} \sum_{i=1}^N \sum_{n=-\infty}^\infty w_i V(\phi_{n,i}).
\end{aligned}
\label{lataction1}
\end{equation}
$V$ is an even polynomial of degree 2 or higher. By convention, the
coefficient of the lowest degree term is set to unity. 

As the lattice gets finer the
matrix $K_{i,j}$ should reproduce accurately more and more eigenvalues of 
the continuum $-\partial_\omega^2$. Define the matrix $Q_{i,j}$ by
\begin{equation}
Q_{i,j} = \begin{cases} w_i K_{i,j}w_j &\mbox{if } i\ne j \\
-\sum_{k\ne i} w_i w_k K_{i,k} &\mbox{if } i=j. \end{cases}
\label{lataction2}
\end{equation}
Let $\mu_k$ be the solutions to the
generalized eigenvalue problem
\begin{equation}
\det_{i,j} [ Q_{i,j}+(1-\mu_k) w_i \delta_{i,j} ]=0.
\label{lataction3}
\end{equation}
Here, $\mu_0=1$ and $\mu_k > \mu_{k+1}>0, \; k=0,1,2...$. 
Then the low eigenvalues of $-\partial_\omega^2$
are approximated by
\begin{equation}
\lambda_k \equiv -\log(\mu_k)/s .
\label{lataction4}
\end{equation}
For well chosen weights, we expect $\lambda_k \approx k(k+1)$ with $k\ge 0$ and
multiplicity $2k+1$. 

The path integral for the partition function is
\begin{equation}
{\cal Z}=\int \prod d\phi_{n,i} e^{-\beta_\omega A[\phi]}.
\end{equation}
One can change variables of integration $\phi_{n,i}\to w_i\phi_{n,i}$. 
This may simplify the form of the action. One cannot forget though that the
weights $w_i$ are essential in correctly 
matching representation of $Q'$ to those of $O(d)$. 
If I put in periodic boundary conditions with $0\le n \le M$, ${\cal Z}=\Tr {\cal T}^M$. ${\cal T}$ is the transfer matrix and it is $M$-independent. It is an integral operator on functions of fixed $n$ fields,
$\phi_j,~j=1,N$. The kernel is symmetric and positive definite:
\begin{eqnarray}
&{\cal D}[\{\phi_i\}]\equiv e^{-\frac12\beta_\omega\sum_{i=1}^N\sum_{j=1}^N w_iw_j \left [ \phi_i K_{i,j} \phi_j\right ] - \frac{\beta_t}{4}\sum_{i=1}^N w_i\phi_i^2-\frac{\kappa}{4}\sum_{i=1}^N w_i  V(\phi_{n,i})}\\&
~\langle \{\phi_i^\prime\} | {\cal T} | \{\phi_i\}\rangle ={\cal D}[\{\phi_i^\prime\}]
e^{\beta_t \sum_{i=1}^N w_i \phi_i^\prime\phi_i}
{\cal D}[\{\phi_i\}]
\end{eqnarray}
The objective is to find the spectrum of ${\cal T}$ and the symmetry properties
of the eigenstates, namely the irreducible representation of the fixed $n$
symmetry $Q'$ and also the quantum number associated with 
the fixed $n$ symmetry which switches simultaneously 
the sign of all fields $\phi_i,\phi_i^\prime$. States even under the latter symmetry
make up the ``even sector'' and states odd under it make up the ``odd sector''. 
For finite $N$, this is a well posed problem.

I now add a technical remark about the evaluation of the
matrix elements of the heat kernel.
Although the heat kernel matrix is evaluated only once 
for a simulation, employing eq.~(\ref{hkm}) might give
negative results at small $s>0$ 
and large separations on the sphere as 
round-off errors accumulate. One might set to
zero entries in the matrix of the quadratic kernel of
the spherical kinetic energy which correspond to 
separations larger than some fixed bound, for example,
require $\hat\omega\cdot \hat\omega^\prime > 0 $. 
Then  one can 
replace the right hand side of eq. (\ref{hkm}) by the leading term in an asymptotic expansion as $s\to 0^+$.
\begin{equation}
<{\hat\omega}|e^{s\partial^2_\omega}|{\hat\omega^\prime}>\sim
\frac{1}{4\pi s} e^{-\frac{d(\hat\omega,\hat\omega^\prime )^2}{4s}}
\sqrt{\frac{d(\hat\omega,\hat\omega^\prime )}{\sin(d(\hat\omega,\hat\omega^\prime ))}}.
\label{asymhk}
\end{equation}
Here
\begin{equation}
d(\hat\omega,\hat\omega^\prime ) =\arcsin[\sqrt{1-(\hat\omega\cdot\hat\omega^\prime)^2}].
\end{equation}
One never has $\hat\omega=\hat\omega^\prime$ but 
$\hat\omega=-\hat\omega^\prime$ does occur unless one puts a bound as above. 
So far, the action is not ultralocal, but local. This is costly
for simulations. One can put more stringent bounds, $\hat\omega\cdot \hat\omega^\prime >\alpha$, where $1>\alpha>0$, 
and even take $\alpha$ to 1 as the lattice is getting refined.
This would produce an ultralocal action. 

Eq.~(\ref{asymhk}) holds for any pair of points $\hat\omega$, 
$\hat\omega^\prime$ for which $\hat\omega\cdot\hat\omega^\prime >-1$, in other words when 
$\hat\omega^\prime$ is not on the cut locus
of  $\hat\omega$. However, Varadhan's asymptotic formula
\begin{equation}
\lim_{s\to 0^+} s \log[<{\hat\omega}|e^{s\partial^2_\omega}|{\hat\omega^\prime}>]=-\frac{1}{4}d(\hat\omega,\hat\omega^\prime )^2,
\end{equation}
holds without restrictions~\cite{hsu}. In practice, 
using this formula
for all $s$ is probably adequate when the number of vertices
is large. By itself, it does not violate $O(d)$ (only the lattice
does) and employing it should not obstruct 
approaching the target theory in the IR as the lattice 
is refined.

\section{Rotational symmetry}

I restrict myself to $d=3$ and $Q=I$, $Q'=I_h$. 
To preserve $I_h$ I need first to determine an appropriate set of vertices. 
I first choose a Cartesian frame in three-space and use the unit
sphere around its origin to label the vertices by points on it, which,
in turn are labelled by unit vectors ${\hat\omega}_i$. Using the 
same frame, the $I$ group elements are labelled by $h$, where the 
$h$'s are
three by three orthogonal matrices of determinant one. $Z_2$ is generated by minus the identity matrix. 

The set ${\hat\omega}_i$ is required to contain only pairs 
$\{ {\hat\omega}_i ,-{\hat\omega}_i\}$ and only complete orbits
under $I$, that is, only sets of the 
form $\{ h{\hat\omega}_i,h \in I \}$. These sets have a number of elements which is a divisor of $I_h$. Including the opposite sign
pairs, the largest orbits have 120 elements each. 
The symmetry requirement
now means that the weights assigned to the vertices are constant
on $Q\times Z_2$ orbits. The action of the elements $g\in I_h$, labelled by orthogonal three by three matrices, 
on the states the transfer matrix acts on is
\begin{equation}
g|\{\phi( {\hat\omega}_i )\}\rangle=|\{\phi(g{\hat\omega}_i)\}\rangle
\end{equation}

The spectrum of ${\cal T}$ will decompose into irreducible 
representation of $I_h$. 
$I$ has 5 irreducible representations~\cite{dressel} 
and $I_h=Z_2\times I$ then obviously has 10. The dimensions of the representations
of $I$ are $1,3,3,4,5$ and the set doubles for $I_h$. 
The irreducible representations of $I_h$ are
labelled $A_g,F_{1g},F_{2g},G_g,H_g, A_u,F_{1u},F_{2u},G_u,H_u$.
Labels with a $g$ subscript are even under $\hat\omega \rightarrow -\hat\omega$ and 
those with a $u$ subscript are odd. $A$ is for singlet, 
$F$ for triplet, $G$ for quadruplet and $H$ for quintuplet.
The decomposition of irreducible representations of $SO(3)$ 
into irreducible representations of $I_h$ can be found in~\cite{gard} (where $T$ is used instead of $F$). 

In the continuum, rotations act on the states 
$|\{\phi({\hat\omega})\}\rangle$ by acting on the 
field argument via the $3\times 3$ matrices and treating $\phi$ as a scalar. 
The irreducible representations are obtained from decomposing
$\phi({\hat\omega})$ into spherical harmonics $\psi({\hat\omega})$. The $\psi$'s are obtained by 
restricting harmonic homogeneous polynomials in the three 
components of 
${\vec \omega}$ to the unit sphere. The representation is identified by the
degree. The degree of homogeneity, $l=0,1,2...$ also determines whether $\psi$ switches sign under $\hat\omega \rightarrow -\hat\omega$ or not. $\psi$'s with even $l$ are invariant
and those with odd $l$ switch sign. 
The dimension of the $l$-th representation is given  by $2l+1$.
They provide representations of $I_h$, not just $I$. 
As representations of $I_h$ they decompose in general into combinations of the 10 irreducible representations of $I_h$. 
The low dimensional irreducible representations of $SO(3)$
corresponding to $l\le 2$ remain irreducible also under $I_h$:
$l=0\rightarrow A_g, ~l=1\rightarrow F_{1u}, ~l=2\rightarrow H_g$.  
Some further cases of interest are $l=3 \rightarrow F_{2u}\oplus G_u$, $l=4 \rightarrow G_g\oplus H_g$ and $l=5 \rightarrow F_{1u}\oplus F_{2u}\oplus H_u$. A singlet of $I_h$ appears for the first 
time in $l=6 \rightarrow A_g\oplus F_{1g}\oplus G_g\oplus H_g$.  

For $1\le l \le 5$ and $|m|\le l$, $\sum_{g\in I_h} \psi_l^m (g{\hat\omega})=0$, where $\hat\omega$ is an arbitrary point on
the sphere. Hence, if the set of vertices consists of complete
orbits, choosing the weights constant on orbits ensures that integrals are 
exactly reproduced by their corresponding 
sums on the 36 dimensional linear space $l\le 5$; this can be achieved
by a 12 vertex orbit on the sphere. 
To push the upper bound on $l$ higher we need
several orbits. Distinct orbits come with distinct weights, which
can be adjusted to zero out discrete counterparts to integrals
of $\psi$'s containing higher $l$'s than 5. For example, using two orbits only, one can zero out the $l=6$ case. The next time
a singlet shows up in the decomposition of a spherical wave is
at $l=10\rightarrow A_g\oplus F_{1g}\oplus F_{2g}\oplus G_g \oplus 2 H_g$. Thus, with two orbits the upper limit on $l$ 
giving exact equality for integrals and sums is pushed to $l=9$.
Evidently, this process can be continued. These facts can be 
learned from~\cite{sobolev}. 

The appearance of singlets in the decompositions reflects the
existence of primitive 
homogeneous polynomials in the three components
of $\omega$ which are invariant under the action of $I_h$. 
They are primitive in the sense that they cannot be expressed 
in terms of other primitives. All $I_h$-invariant polynomials are
polynomials in three primitives of degrees 2,6,10~\cite{flato}. 
The first is an invariant of $O(d)$ and restricts
to a constant on the sphere. The next two 
primitive polynomials associated with $I_h$ are primary objects in the process of understanding  how full $SO(3)$ is violated. 

The group $I_h$ is generated by reflections in 3 planes through
the origin of three space. $I_h$ contains more pure
reflections, corresponding to 15 mirror 
symmetry planes in total. 
All finite groups of this type are classified~\cite{coxeter}. One good place to learn the subject from is~\cite{humphreys}. Specifically focused on the Icosahedron  is the classic~\cite{klein}. 

Particle Physicists are more familiar with  crystallographic Coxeter groups because of their
connection to the representation theory of Lie Algebras. Exceptional Weyl groups in this category have
already been exploited for Particle Physics related problems 
as they provide enhanced rotational
invariance in specific dimensions. Specific Particle Physics applications can be found in~\cite{f4}. Radial quantization does not require a crystallographic group. There are few non-crystallographic groups and the ones of interest are denoted by
$H_3$ and $H_4$ respectively. $H_3$ is $I_h$ and has 15 reflections, as mentioned already. $H_4$ has 60 reflections and
$120^2$ elements. $H_3$ and $H_4$ provide enhanced rotational 
invariance in dimensions 3 and 4 respectively.

The structure of a reflection group is quite simple geometrically. One has on the sphere a fundamental region bounded by 3 basic planes. The group
acts on this fundamental region producing 
new ones and tessellates 
the sphere by $2|Q|$ 
such spherical triangles. The entire 
spheres gets fully covered exactly once. 
One can pick one fundamental region, 
label it as the unit element of the group and label all its
images by the connecting group element. The covering is bipartite, 
according to the $Z_2$ factor in the reflection group. 
Once a fundamental region is chosen, any point inside it 
(that is not on any boundary component) has an orbit consisting
of a number of points equal to the group order. Points
on the boundary will generate orbits of lower multiplicity.
All multiplicities are divisors of the number of group elements.
From the point of view of ``vertex economy'' one likes smallish 
orbits since they allow 
an independent weight parameter with whose help one can 
zero out more and more $SO(3)$ irreducible representations in 
the cubature formula. It is not clear that the principle of
``vertex economy'' really needs to be taken seriously when
designing a large scale simulation, but it certainly is
useful in finding easily manageable test cases. 

The spectrum of the transfer matrix will decompose into 
many copies of each of the 10 irreducible representations.
In general, it will be difficult to disentangle this structure
for many reasons. This should be substantially easier 
close to the bottom of the spectrum, where the lowest 
dimension scaling fields have their corresponding states. 
Numerical simulations cannot access regions of high energy states anyhow. 

\section{Simple examples}

The aim of this section is to investigate simple cases where
various ingredients of the cubature approach can be tested.

\subsection{Spectrum of quadratic kernel}

One criterion to determine how well eq. (\ref{lataction1}) works 
is to work out the spectrum as described by eqs. (\ref{lataction2}), (\ref{lataction3}), (\ref{lataction4}) on coarse lattices. 

\begin{figure}
\includegraphics[width=0.4\textwidth]{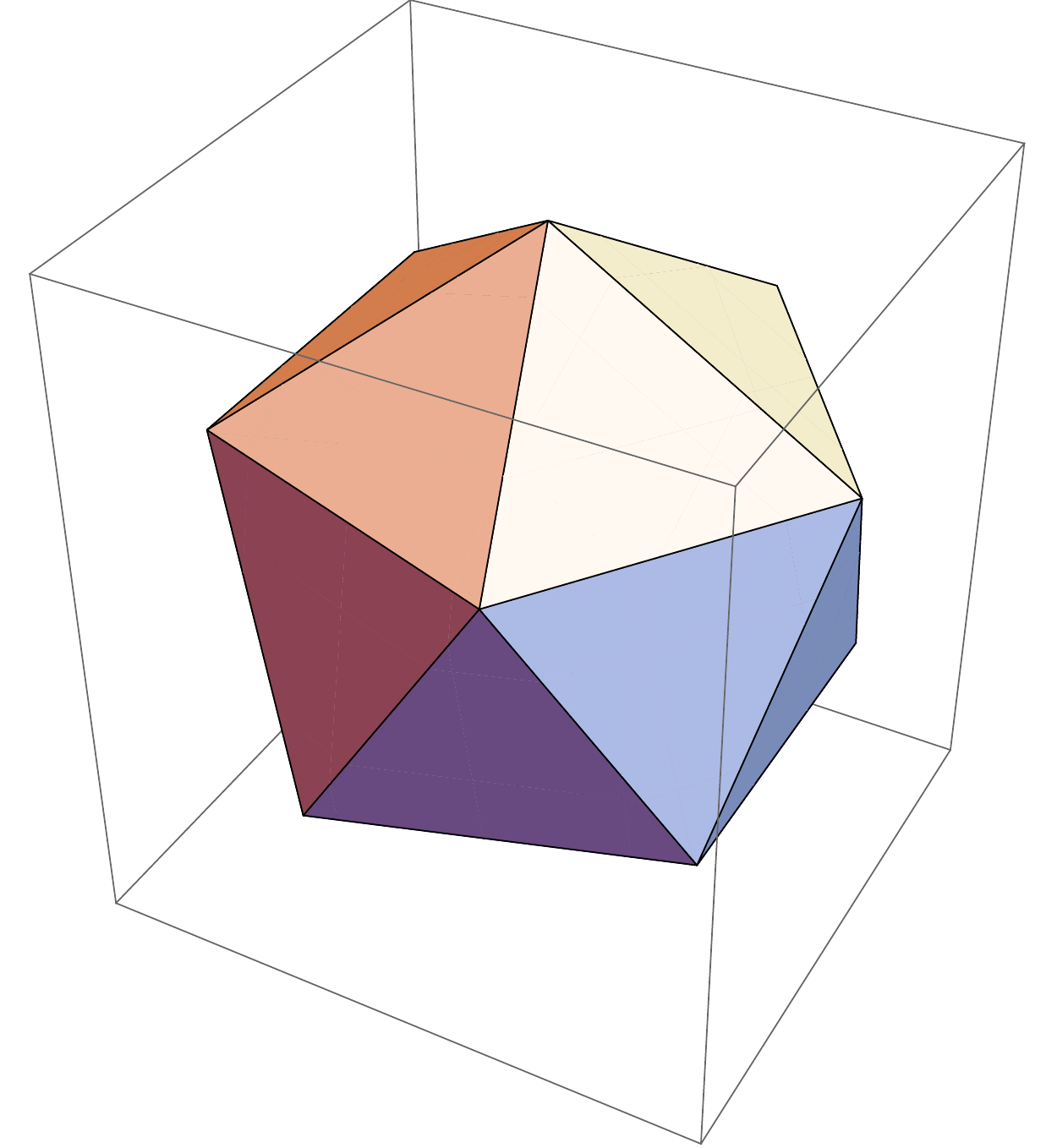}
\caption{Icosahedron}
\label{ico}
\end{figure}

The coarsest lattice on the two sphere I consider 
consists of the corners of 
a regular Icosahedron in Figure.~\ref{ico}. This lattice has 12 vertices. The 120 
elements of $I_h$ 
permute the vertices in various ways. The matrix $K_{ij}$ 
is invariant under conjugation by elements of $I_h$ acting on 
the vertices. $K$ acts on a 12 dimensional space. This space
decomposes into $A_g\oplus F_{1u}\oplus F_{2u}\oplus H_g$.

Taking $s=0.7$ for the spherical heat kernel and $w_1=1/12$ ~I found for the $\lambda_k$ of eq.~(\ref{lataction4}) the following values: $\lambda_0=0.,\lambda_1=2.,\lambda_2=5.9999,
\lambda_3=10.7896$, with multiplicities $1,3,5,3$ respectively.
Thus, the $l=0,1,2$ eigenvalues and multiplicities are well
reproduced, but $l=3$ not. With only 12 dimensions 
available there are not enough states to provide for 
a full set of states descending  
from the full $l=3$ multiplet. The $l=3$ multiplet is expected to split, and the quadruplet is missing. 
It is not possible to estimate the split.
Nevertheless, the numerical value of the eigenvalue is not outrageously far from the correct value of 12. 
We see that even a small orbit does as good a job as one might reasonably expect in terms of what can be read off the action.

\begin{figure}
\includegraphics[width=0.4\textwidth]{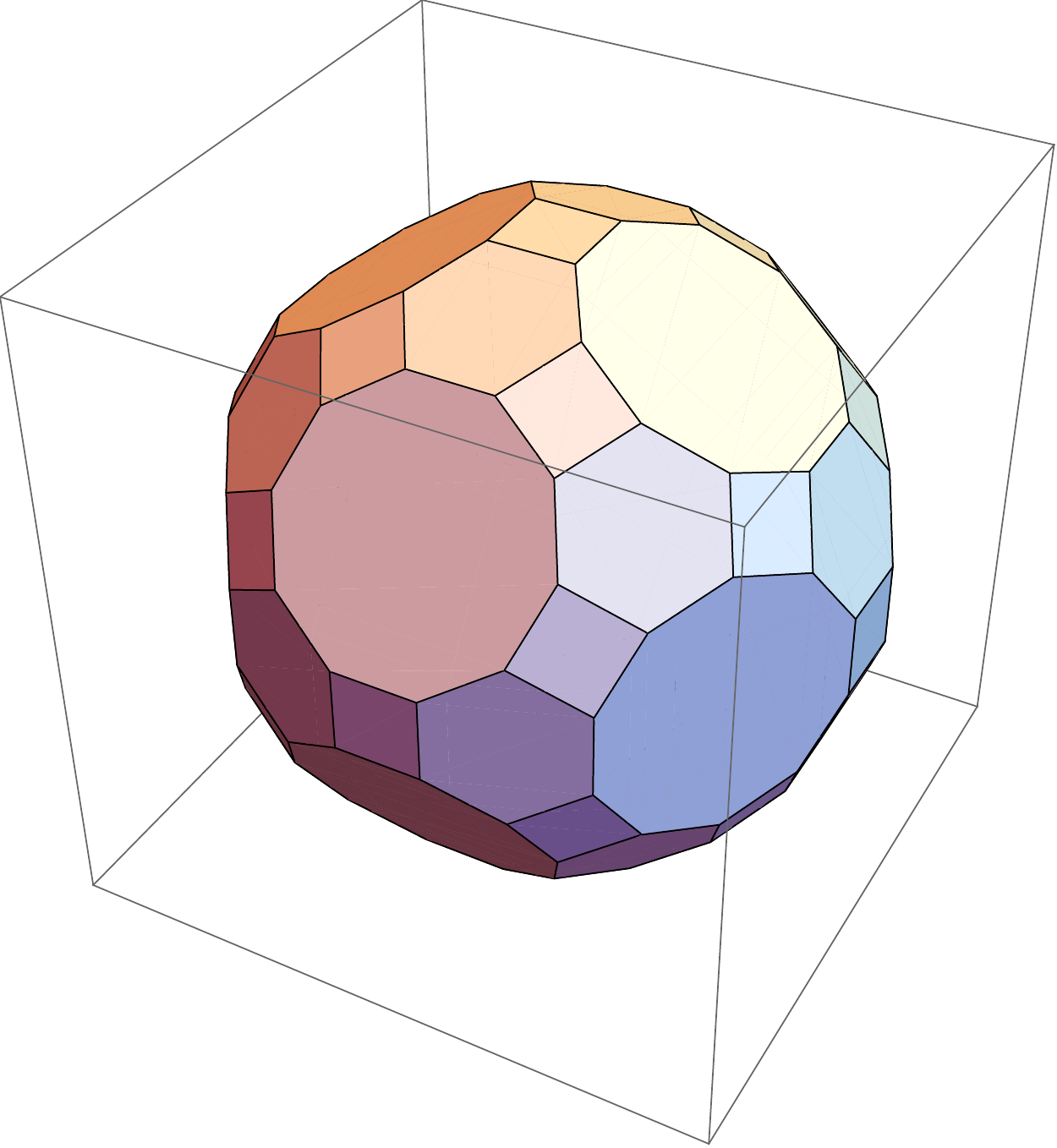}
\caption{Great Rhombicosidodecahedron}
\label{greatrh}
\end{figure}

Having seen how a very small orbit performs, I turn to a maximally large orbit, taking a lattice with 120 vertices.
I choose the Great Rhombicosidodecahedron of 
Figure~\ref{greatrh} for this purpose.
Each vertex, $X$, lies at the intersection of the spherical bisectors 
of the spherical triangles making up the fundamental region. The fundamental region can be constructed by adding the face centers
and edge centers to the Icosahedral spherical 
tessellation. Any spherical triangle
with 3 vertices in the same triangular Icosahedral face, 
consisting of one face vertex, one center vertex and one 
edge center, makes up a fundamental region. The Icosahedron has 20 faces and each has 6 fundamental regions. Therefore there
are 120 X-type vertices. If connected by segments of
great circles perpendicular to the edges of the fundamental regions, all segments are of equal length. For 
gaining some familiarity with these constructions I recommend~\cite{cromwell}.
A glance at Figure~\ref{greatrh} shows that the vertices are not distributed in a very uniform manner. It remains to be seen to what extent this impression correctly
reflects on the usefulness of this lattice. 

In this case the representation of $I_h$ is the regular one, that is, 
each of the 10 irreducible representations enters into 
the decomposition a number of times equal to its dimension. Looking at the eigenvalues $\lambda_k$ as before, I can go higher up in level.
For the $l$-levels that split under $I_h$ I take
an average over the various $I_h$ irreducible representations 
that contribute, weighted by their sizes. This gives
me numbers to compare to the $l(l+1)$ values. I also take the
spread of the $\lambda_k$ making up the contributions as a measure of the split, $S_l$. In this way I found:
$\lambda_{l=0}=0$, $\lambda_{l=1}=2$, $\lambda_{l=2}=5.9999$, $\lambda_{l=3}=12.0466$,  $\lambda_{l=4}=20.1072$, $\lambda_{l=5}=30.2122$, $\lambda_{l=6}=42.8426$, with $S_l=0$ for $l=0,1,2$ and $S_3=0.7507$, $S_4=0.6882$, $S_5=1.3273,1.0562$, $S_6=9.2407,9.0609$. Where two numbers appear, I took various
combinations of the multiplets into which the particular $l$ decomposed to get some feel. Looking at $l=6$ it is clear that 
the level identification looses meaning after $l=5$. Nevertheless, even for $l=6$, the  average number is close to $l(l+1)$. Multiplicities are always $2l+1$. Looking at the $l=3$
case, we see a much better match with the expected value of 
$12$ then before. The weighted average is much closer to
the expected value than the split would indicate. This looks
like an effect of symmetry breaking dominated by a term
in an expansion at first order. Perhaps the split of the level
$l=3$ seen in ~\cite{ourplb} is of similar origin. That is,
in the continuum limit the eigenvalues associated with the
icosahedral arrangement of conic singularities in an otherwise
flat manifold differs from the spherical, fully rotationally
symmetric essentially by a first order perturbation in a symmetry
breaking term. This term ought to be predominantly
proportional to the primitive $I_h$-invariant of order 6.

So far I have only looked at single orbits where the weight is fixed to be the inverse of the orbit size. It is obvious that
beyond $l=2$, as expected, splits will occur and that 
they have a structure that looks perturbative. 

More
precisely, if I imagined writing a
continuum ``effective theory'' description of the
discrete approximation, the continuum $K$ would have corrections which would be still continuum kernels, but
break $O(3)$ to $I_h$. I could order these corrections
by looking at the action restricted to field sectors 
spanned by low $l$ spherical degrees. The leading correction would be one that becomes felt at the lowest $l$. The corrections, when sandwiched between the states
with lowest $l$'s which are going to split must generate
an invariant under $I_h$ which is not an invariant under
$O(3)$. The lowest degree of that integrand is 6, as 
already discussed. 
The affected $l$'s of the fields would than have to be, 
as expected, $l=3$. There is only one parameter that enters, associated with the degree 6 invariant. 
To leading order the effect cancels out in
the multiplicity weighted average. That makes the 
deviation of $\lambda_{l=3}$ from 12 quite small.
At $l=5$ the degree 10 invariant enters and a larger split
occurs, reflecting the extra coupling. 
I plan to report separately on a more detailed analysis of
these breaking effects~\cite{future1}.

\begin{figure}
\includegraphics[width=0.4\textwidth]{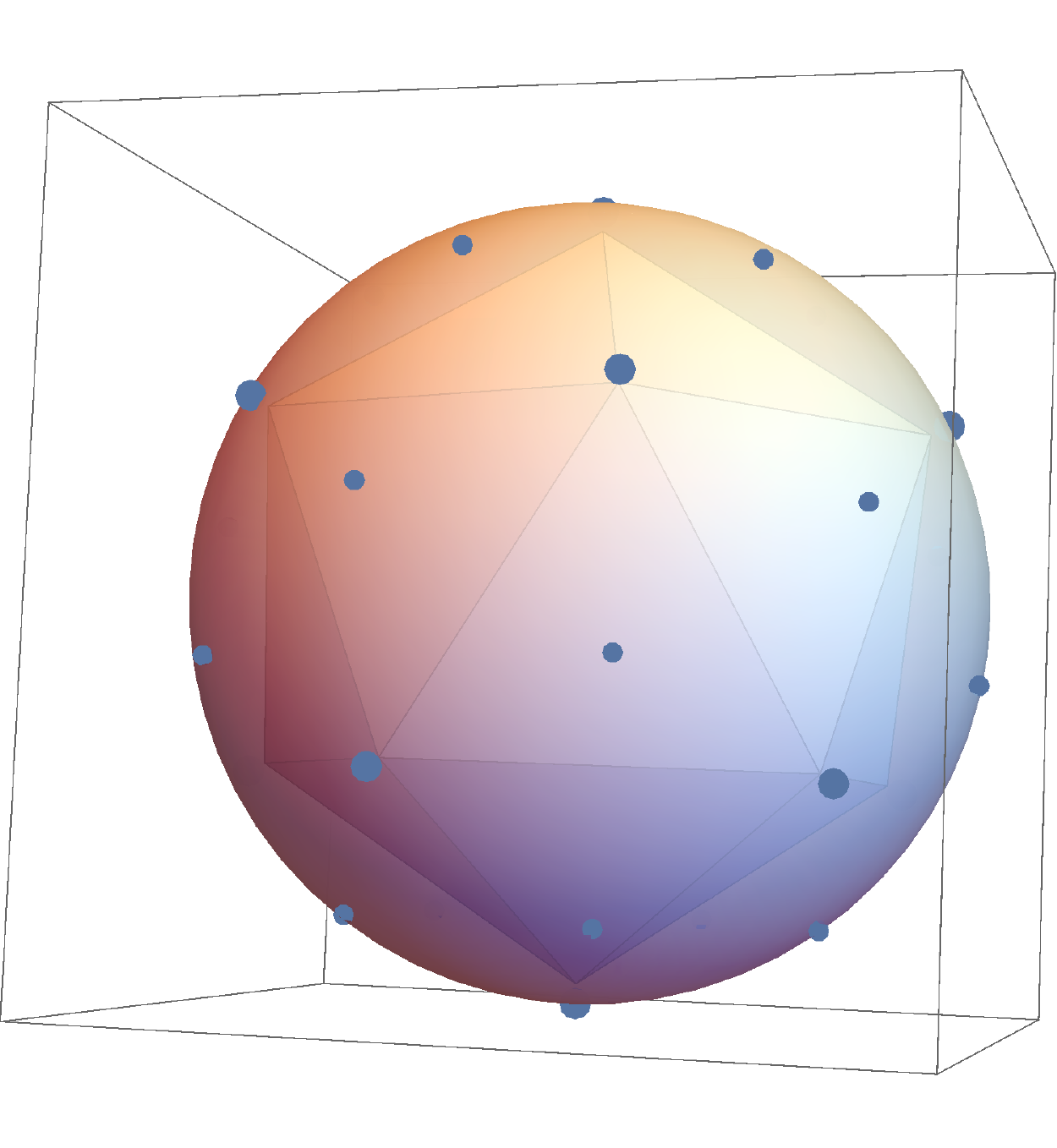}
\caption{A 32 vertex, 2 orbits arrangement.}
\label{ico_w_ctrs}
\end{figure}

Now I wish to look at a minimal example in which I have two orbits, and, by adjusting their weights I can eliminate the split
in the $l=3$ level. I take the Icosahedron and add to it the centers of the faces. This gives me in total 32 vertices and
2 orbits. There is one free parameter, which is the ratio of
the two weights. See Figure~\ref{ico_w_ctrs}. 
I want to use it in order to zero out the split.
At the level of cubature this is a well known problem, solved long ago~\cite{sobolev}. I require that the weights be
such that at the level of simple cubature, where the action 
density (not the fields) is expanded in spherical harmonics, there be exact agreement between the sum and integral for $l=6$. In fact, I am zeroing out the
coupling of the degree 6 invariant. 
As explained before, this ensures sum and integral agreement
of the simple cubature formula all the way up to $l=9$. 
If $s$ is large enough, as I explained, one expects no
breaking effects up to and including level $l=4$. That is,
both the splits of $l=3$ and $l=4$ should be small.

Picking the weights $w_c=\frac{f}{20(1+f)}$ and
$w_v=\frac{1}{12(1+f)}$ for the orbit of triangle centers and that of original vertices respectively, and using $f=1.8$, the
exact value derived from the simple cubature equation, I obtained, 
again with $s=0.7$, $\lambda_0=1$, $\lambda_1=2$, $\lambda_2=6$, 
$\lambda_3=12$, $\lambda_4=20$ with increasing deviations as $l$ increases. The largest deviations are of order $10^{-7}$, which
is the order of the split at $l=4$. As $s$ is increased the
splits drop dramatically even further, in accordance with our analysis earlier.  At $l=5$ the structure has totally deteriorated: indeed
up to $l=4$, 25 states were accounted for. The $l=5$ state
would add 11 states, but we only have 32-25=7 left. These 7 states come in two triplets and a singlet. 

The findings so far support the approach, but only at the level
of the quadratic part of the action, which, from the point of
view of field theory corresponds to free field theory. I
need to get some feel for the situation in an interacting
situation. 

\subsection{Large N}

Consider the CFT generated by the continuum linear or nonlinear $O(N)$ model in
three Euclidean dimensions. It is well known that at leading order in $1/N$ the model can be described by a free massless
field theory for $N$ scalar fields. One needs to adjust one
coupling to make the theory massless, and like any free massless
scalar theory it is also a CFT. 

The radially quantization of massless scalar fields and
the role of conformal invariance were exposed in~\cite{jackiw}.~\cite{jackiw} 
derives the radially quantized
version of the field theory from the same model traditionally quantized on flat
two dimensional subspaces of $R^3$. 
In addition to changing variables the correct cylinder structure
$R\times S_2$ will hold for the case that the scalar fields
are rescaled by the appropriate power of the radius. In three
dimensions this has the effect of replacing $l(l+1)$ by $l(l+1)+1/4=(l+1/2)^2$. 
The dimensions come from taking square roots
of this factor. The 1/4 is crucial in order to get 
infinite equally spaced towers in the spectrum. 
In his paper, Cardy~\cite{cardy}, also deals with the
$O(N)$ model and shows that the 
extra $1/4$ is equivalent to
the condition for criticality in flat space, obtained by
solving the gap equation for the massless case. He does this by
endowing the sphere with a radius $R$, and matching to
flat space at large $R$. Another way to
get the 1/4 is to postulate conformal coupling of the
scalar to the round metric on the two-sphere. 

In any case, we see in this example explicitly how the flat space
adjustment needed for criticality is equivalent to the
requirement of having states in the odd $Z_2$ sector organize themselves into equally spaced towers. 

Working out the explicit 
CFT structure to higher orders in $1/N$ 
rapidly becomes a complicated problem~\cite{ruhl}.

In~\cite{ourplb} we showed that in two dimension the $O(N)$ model
does not admit an adjustment which would make the towers 
equally spaced. This is consistent with the model having to break scaling at the quantum level.

\subsection{A transfer matrix example}

So far I have checked that the construction of an action 
thinking in terms of cubature formulas has a chance to work.
Quantum mechanically however, all that was checked was free field
theory. 

Now I want to work out one example which is fully interacting.
I want an example that I can do almost analytically. By this
I mean that I can, in a matter of a few minutes on
the computer, get very high accuracy
results without using anything stochastic. 

The example consists of the Ising model defined on the
sphere with 12 Icosahedral vertices. The associated transfer 
matrix 
is $2^{12} \times 2^{12}$ and can be fully diagonalized
with standard routines. I am forced to use the Ising model in order to minimize the number of values the fields can take, while
still having a global internal $Z_2$ symmetry. 

Since the heat kernel formalism has been checked already, I 
am not bound to it. I only adopt the idea to use a non-local 
interaction and am going to adjust it the best I can. 

It is well known that the Ising model can be written in terms of
continuum fields~\cite{zinn}. 
This makes the application of mean field theory
straightforward. I do not need the explicit expressions.
The main point is that there is a quadratic kernel whose eigenvalues and behaviour under rotations are still relevant
although the original fields were discretely valued. 
However, the potential is not polynomial and therefore 
the symmetry analysis I presented before does not apply.
There are also other problems, making a transfer matrix 
in terms of the continuous fields untenable. 

The strategy is as follows: first treat the quadratic part as if
this was a free theory with a continuously valued real scalar field. Adjust in such a manner that it give the best rendition
of the spherical Laplacian possible. Then use it to define
the spin-spin interaction in the Ising model. The hope is that
this structure would ensure that one can find a large region of
parameter space where the order of low
states is what one expects from the model. Next, introduce 
two more couplings, and search for a pseudo critical point.
This point is characterized by some CFT spectral regularity
holding there at a reasonable level of accuracy. 
The criterion is that spacings between the two lowest 
rungs in tentatively identified towers agree between the
odd and even sector. I also require to tentatively identify 
the state corresponding to the energy momentum tensor.
If all three independent determinations of scale
agree with other, I can get rough numbers for the
dimensions of some of the lowest primaries. The main
intention is to see the right structure and 
numbers in the right ballpark. 
It would be unrealistic to expect more from
such a small system. 
I diagonalize the transfer matrix to get its spectrum
and symmetry properties of the lowest states. 

\subsubsection{The spin-spin interaction}

\begin{figure}
\includegraphics[width=0.3\textwidth]{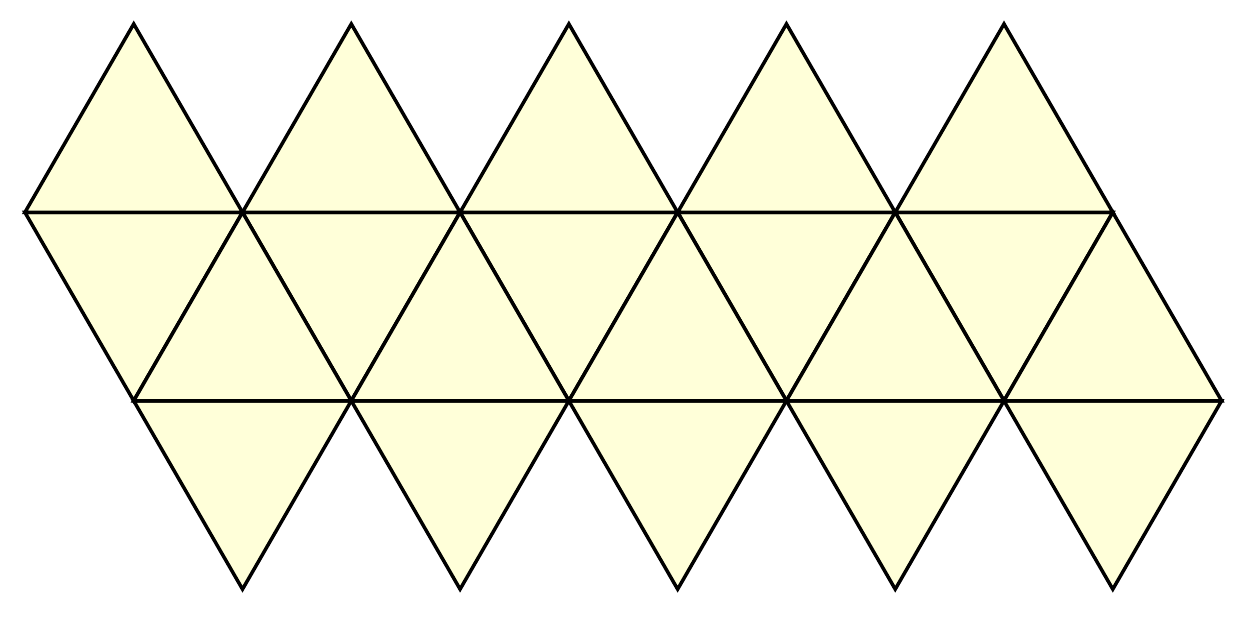}
\caption{Icosahedral Net}
\label{iconet}
\end{figure}

On each site of one spherical shell we have a spin $\sigma_i=\pm 1$. The sites are labelled by 1 to 12. The labels, according
to the Icosahedral net in Figure~\ref{iconet} go as follows:
the top row are all site 1. The next horizontal row has
labels 2,3,4,5,6,2 left to right, followed by labels
7,8,9,10,11,7 and the bottom row are all site 12. 

The term in the action I am now focusing on is
\begin{equation}
{\cal A}=\sum_{i\ne j} \sigma_i C_{i,j}\sigma_j
\end{equation}
My objective is to determine the 
off diagonal entries in the symmetric
matrix $C_{i,j}$. I do not want this choice, in itself, to 
violate $O(3)$. So, $C_{i,j}$ only depends on $\hat\omega_i
\cdot\hat\omega_j$. The distances between non-identical 
sites take only three values. If I only made $C_{i,j}\ne 0$ 
zero for the shortest non-zero distance, there would be no
indication that the sites reside on a sphere rather than
on the corners of a solid Icosahedron. The 12 dimensional representation of $I_h$ provided by this set of sites 
decomposes into $A_g\oplus F_{1u}\oplus F_{2u}\oplus H_g$, 
as already mentioned. I can introduce 3 different parameters 
corresponding to the 3 values of distances; they correspond to
the 3 non-trivial irreducible representations above.

With the above labelling the most general $C_{i,j}$ 
$O(3)$-invariant matrix is
\begin{equation}
C=\left(
{\scriptscriptstyle\begin{array}{cccccccccccc}
 0 & a & a & a & a & a & b & b & b & b & b & c \\
 a & 0 & a & b & b & a & a & b & c & b & a & b \\
 a & a & 0 & a & b & b & a & a & b & c & b & b \\
 a & b & a & 0 & a & b & b & a & a & b & c & b \\
 a & b & b & a & 0 & a & c & b & a & a & b & b \\
 a & a & b & b & a & 0 & b & c & b & a & a & b \\
 b & a & a & b & c & b & 0 & a & b & b & a & a \\
 b & b & a & a & b & c & a & 0 & a & b & b & a \\
 b & c & b & a & a & b & b & a & 0 & a & b & a \\
 b & b & c & b & a & a & b & b & a & 0 & a & a \\
 b & a & b & c & b & a & a & b & b & a & 0 & a \\
 c & b & b & b & b & b & a & a & a & a & a & 0 \\
\end{array}}
\right)
\end{equation}

One has
\begin{equation}
C=aX+bY+cZ,
\end{equation}
and the matrices $X,Y,Z$ commute. 

Using the explicit labelling
it is easy to verify that the permutation matrix $Z$ is the inversion. 
\begin{equation}
Z=\left(
{\scriptscriptstyle\begin{array}{cccccccccccc}
 1 & 0 & 0 & 0 & 0 & 0 & 0 & 0 & 0 & 0 & 0 & 0 \\
 0 & 0 & 1 & 0 & 0 & 0 & 0 & 0 & 0 & 0 & 0 & 0 \\
 0 & 0 & 0 & 1 & 0 & 0 & 0 & 0 & 0 & 0 & 0 & 0 \\
 0 & 0 & 0 & 0 & 1 & 0 & 0 & 0 & 0 & 0 & 0 & 0 \\
 0 & 0 & 0 & 0 & 0 & 1 & 0 & 0 & 0 & 0 & 0 & 0 \\
 0 & 1 & 0 & 0 & 0 & 0 & 0 & 0 & 0 & 0 & 0 & 0 \\
 0 & 0 & 0 & 0 & 0 & 0 & 0 & 1 & 0 & 0 & 0 & 0 \\
 0 & 0 & 0 & 0 & 0 & 0 & 0 & 0 & 1 & 0 & 0 & 0 \\
 0 & 0 & 0 & 0 & 0 & 0 & 0 & 0 & 0 & 1 & 0 & 0 \\
 0 & 0 & 0 & 0 & 0 & 0 & 0 & 0 & 0 & 0 & 1 & 0 \\
 0 & 0 & 0 & 0 & 0 & 0 & 1 & 0 & 0 & 0 & 0 & 0 \\
 0 & 0 & 0 & 0 & 0 & 0 & 0 & 0 & 0 & 0 & 0 & 1 \\
\end{array}}
\right)
\end{equation}

Projecting
on the two subspaces invariant under $Z$, $X$ decomposes as $X=X_1+X_2$ and $Y$ decomposes as $Y=Y_1+Y_2$. It is easy to
check that $ X_1=Y_1, X_2=-Y_2$. Hence we can 
choose $X_1 , X_2 ,Z$ 
as the complete set of commuting operators for this problem.

Both $X_1$ and $X_2$ have 6 dimensional kernels. The non-zero
spectrum of $X_1$ consists of $5$ (singlet) and $-1$ (quintuplet).
The corresponding states are +1 eigenvectors of $Z$. 
The non-zero spectrum of $X_2$ consists of $\pm\sqrt{5}$, 
two triplets. The corresponding states are -1 eigenvectors of
$Z$. So, from $X_1$ we have identified the states
in $A_g\oplus H_g$ and from $X_2$ the states in 
$F_{1u}\oplus F_{2u}$. In order to identify 
which of the $\pm \sqrt{5}$ triplets is $F_{1u}$ and which is
$F_{2u}$ we need to look at the action of a rotation by
$72^\circ$ about an axis of symmetry of the Icosahedron. 
For the axis connecting the top to bottom vertices in 
Fig.~\ref{iconet} the action leaves fixed 
vertex 1 and vertex 12
and cyclically permutes by one step the two remaining horizontal rows simultaneously. The corresponding matrix, $G$ is
\begin{equation}
G=\left(
\begin{array}{cccccccccccc}
 1 & 0 & 0 & 0 & 0 & 0 & 0 & 0 & 0 & 0 & 0 & 0 \\
 0 & 0 & 1 & 0 & 0 & 0 & 0 & 0 & 0 & 0 & 0 & 0 \\
 0 & 0 & 0 & 1 & 0 & 0 & 0 & 0 & 0 & 0 & 0 & 0 \\
 0 & 0 & 0 & 0 & 1 & 0 & 0 & 0 & 0 & 0 & 0 & 0 \\
 0 & 0 & 0 & 0 & 0 & 1 & 0 & 0 & 0 & 0 & 0 & 0 \\
 0 & 1 & 0 & 0 & 0 & 0 & 0 & 0 & 0 & 0 & 0 & 0 \\
 0 & 0 & 0 & 0 & 0 & 0 & 0 & 1 & 0 & 0 & 0 & 0 \\
 0 & 0 & 0 & 0 & 0 & 0 & 0 & 0 & 1 & 0 & 0 & 0 \\
 0 & 0 & 0 & 0 & 0 & 0 & 0 & 0 & 0 & 1 & 0 & 0 \\
 0 & 0 & 0 & 0 & 0 & 0 & 0 & 0 & 0 & 0 & 1 & 0 \\
 0 & 0 & 0 & 0 & 0 & 0 & 1 & 0 & 0 & 0 & 0 & 0 \\
 0 & 0 & 0 & 0 & 0 & 0 & 0 & 0 & 0 & 0 & 0 & 1 \\
\end{array}
\right).
\end{equation}
Now, using the character table and the matrices $G$ and $G^2$, one determines that
the $\sqrt{5}$ eigenvalue corresponds to $F_{1u}$ 
and the $-\sqrt{5}$ eigenvalue corresponds to $F_{2u}$.
Hence the $-\sqrt{5}$ $X_2$ eigenspace should be thought
of as descending from $l=3$. 

The spectrum of $C$ is linear in the parameters $a,b,c$
and we know now to which continuum $l$ each invariant space should be assigned. It is convenient to add a new variable, $x$, which provides an overall shift 
of the spectrum of $C$. Now, we have just enough
freedom to ensure that the spectrum of the matrix 
provides eigenvalues associated with $l=3,2,1,0$ 
given by $-l(l+1)$ in ascending order $-12,-6,-2,0$.
I end up with 
\begin{equation}
C=- 6 \; {\mathbb{I}}+\tau X +(1-\tau) Y +Z,
\label{C}
\end{equation}
which has a spectrum as close as possible to the continuum Laplacian.
Here $\tau=\frac{1+\sqrt{5}}{2}$, the golden ratio.
This equation would be directly 
relevant to continuously valued
fields. In the Ising case the contribution of the identity
matrix is irrelevant. One will have to rescale the matrix
$C$ by a coupling $\beta_x$. 
Because of the relationship
to an action in terms of a continuum field I already mentioned, 
all one can expect is to get the right order and relative magnitudes of eigenvalues. In other words, an overall scale
for the energies will have to be determined from the
results. This is something one expects. The purpose of the
entire exercise is to find a form of the spin-spin
interaction that has some likelihood of producing at least the right ordering of states in the discretized version. Here ``right'' is
with respect to the group theoretical identification of
the various multiplets with their continuum ``parents''.

\subsubsection{The spectrum of the transfer matrix}

The transfer matrix acts on a 4096 dimensional Hilbert
space. A distinct fixed time slice spin configuration 
$\{s_j\}$ labels each element in a basis. The numbering
is the same as above, based on the Icosahedral net. 
The spins
$\sigma_j$ are given by $\sigma_j=2 s_j-3$, so $s_j=1,2$.
Any basis element can be labelled by $J=s_1+\sum_{j=2}^{12} 
(s_j-1)2^{j-1}$, where $J=1,...,4096$. The spin at
vertex $i$ in configuration $J$ is denoted by
$\sigma_i(J)$. The inverse map is denoted by 
$J(\{\sigma_i\})$.

I define diagonal matrices ${\cal D}$:
\begin{equation}
{\cal D}_{J,K}=e^{\frac12 \beta_x \sum_{1\le i\ne j\le 12}
\sigma_i(J) C_{i,j} \sigma_j(K)}\delta_{J,K},
\end{equation}
with $C$ given by eq. (\ref{C}). 
The transfer matrix is given by:
\begin{equation}
{\cal T}_{J,K} = {\cal D}_{J,J} 
e^{\beta_t\sum_{i=1}^{12} \sigma_i(J)\sigma_i(K)}
{\cal D}_{K,K}.
\end{equation}

The internal global $Z_2$ symmetry acts by $\sigma_i 
\rightarrow -\sigma_i,\; \forall i$, which defines the action on 
the configurations $J,K..\;$. Viewing the $(s_i-1)$ as bits, 
the action is by two's complement on the integer labelling the configuration. Symmetrizing and anti-symmetrizing ${\cal T}$ with respect to this $Z_2$ 
gives 2 matrices of size $2048\times 2048$ each, 
${\cal T}^A$ and ${\cal T}^S$ . 
Similarly, the action of the inversion $Z$ can be
mapped into an action on the labels $I,J..\;$. I
can then decompose ${\cal T}^A$ and ${\cal T}^S$ 
separately by projecting on $\pm 1$ eigenspaces of inversion.

Next the four $2048\times 2048$ matrices
are numerically diagonalized. That takes little time.
I collect the highest eigenvalues of all four matrices. I then know multiplicities, the internal $Z_2$ sector 
and whether the states switch sign under inversion
or not. It is not necessary to calculate the eigenvectors
for this. 

By analogy with~\cite{ourplb}, I have set $\beta_x=0.160$
and varied $\beta_t$ in the search for a roughly
consistent scale determination. I now present a ``good'' case as far as I can tell after searching not very exhaustively. The logarithm of the highest state energy is 9.8907388587432123. This state is the vacuum. It resides in the even sector where it belongs. I shall subtract from it the logarithms of all the lower energy states. These are the excitation energies. Up to a
common rescaling they should provide dimensions of 
primaries and descendants. 
Below are numerical results and tentative interpretations of states for $\beta_t=0.225$. 
I use standard notation for the states.
\vskip 0.5cm 
\begin{tabular}{|l|l|l|p{6cm}}
\hline
Sect. & Mult. &Excitation Energy \\ \hline
Even &1 & 1.3142776638306035     \\ \hline
Even &3 & 2.1649446841074473     \\ \hline
Even &5 & 2.3932548321963418     \\ \hline
Even &5 & 2.7868410430944044     \\ \hline
Even &4 & 2.8973684611035200     \\ \hline
Even &3 & 2.9092189893100091     \\ \hline
Even &5 & 3.0638226605671877     \\ \hline
Even &1 & 3.1255791565017601     \\ \hline
Even &4 & 3.5269178230913276     \\ \hline
\end{tabular} 
\quad
\begin{tabular}{|l|l|l|l|p{6cm}}
\hline
Sect. & Symbol & $l$ & Energy \\ \hline
Even   &$\epsilon$ &0   &1.3142776638306035 \\ \hline     
Even   &$\epsilon$ &1   &2.1649446841074473 \\ \hline
Even   &$T_{\mu\nu}$ &2  &2.3932548321963418 \\ \hline
Even   & $\epsilon$ &2   &2.7868410430944044 \\ \hline
Even   &skip 12 states  && \\ \hline
Even   &$\epsilon^\prime$ &0 &3.1255791565017601 \\ \hline     
\end{tabular}
\vskip 0.5cm
\begin{tabular}{|l|l|l|p{6cm}}
\hline
Sect. & Mult. &Excitation Energy \\ \hline
Odd &1 &0.42744457201564146 \\ \hline    
Odd &3 &1.2689723266444659   \\ \hline  
Odd &5 &1.9039503886862708     \\ \hline
Odd &3 &2.3895288048861216    \\ \hline 
Odd &1 &2.4721280980424005    \\ \hline 
Odd &3 &3.2980151032493685     \\ \hline
Odd &5 &3.5915520069173139     \\ \hline
Odd &4 &3.6018838120548367     \\ \hline
Odd &3 &3.7001232625160938     \\ \hline
Odd &5 &3.7286936205459451     \\ \hline
\end{tabular}
\quad
\begin{tabular}{|l|l|l|l|p{6cm}}
\hline
Sect & Symbol & $l$ & Energy \\ \hline
Odd &$\sigma$  & 0 & 0.42744457201564146 \\ \hline 
Odd &$\sigma$ & 1 &1.2689723266444659   \\ \hline
Odd &$\sigma$ & 2  &1.9039503886862708 \\ \hline
Odd &$\sigma$ & 3 ($F_{1u}$) &2.3895288048861216 \\ \hline 
Odd &$\sigma^\prime$ & 0 & 2.4721280980424005  \\ \hline
\end{tabular}
\vskip 0.5cm
Determining the scale from the spacings of the lowest
tower members in the even and odd sector I get
$\rho=1.182(6)$ and this produces the 
following dimensions after division by $\rho$:  $\dim(\epsilon)=1.553$, $\dim(\epsilon^\prime)=3.694$, $\dim(T_{\mu\nu})=2.829$, $\dim(\sigma)=0.505$ and $\dim(\sigma^\prime)=2.922$. 
I cannot make any serious claims about the validity (to
say nothing about the accuracy) of these numbers.
My point is that they are in the right ball park. 
There are serious numerical estimates in the literature
to compare to~\cite{vicari}: $\dim(\epsilon)=1.41$, 
$\dim(\epsilon^\prime )=3.8$, $\dim(T_{\mu\nu})=3$, 
$\dim(\sigma)=0.518$, $\dim(\sigma^\prime) \sim 4.5$. 
It is
encouraging to see that the energy momentum
tensor dimension comes out quite close to 3. 
One also sees how the breakup of higher $l$ 
multiplets mixes up the order of states pretty early.
It would be unrealistic to expect anything more from a spherical shell approximated by just 12
points. 

\section{Summary}

This paper has led me to a quite flexible procedure to
set up a Monte Carlo simulation of the radially quantized $\phi^4$ model in three dimensions. I have sketched how
an analysis would have to be executed. Quite a few details have been left open and adjustments would need to be 
done as more experience is being accumulated. 
In broad lines, the procedure is
\begin{itemize}
\item Define a sequence of spherical lattices with a convenient decomposition into orbits.
\item Determine a set of weights, one per orbit, all 
positive, such that splits of $l$-mulitplets in the 
kinetic energy quadratic form are eliminated to a
highest possible level $l=l_{\max}$.
\item Define the action of a $\phi^4~3\, d$ theory
using the weights and the heat kernel kinetic energy.
\item Experiment with the choice of $s$ and various
approximations to the heat kernel function.
\item Make initial Monte Carlo
runs to locate consistent level spacings, once 
it is determined that the actual multiplets hold together
well (small splits for a number of $l$'s increasing with
the number of orbits). Eliminating splits in the
quadratic form of the kinetic energy will not
exactly eliminate splits among the actual eigenstates
of the transfer matrix, as inferred from various two point
correlations.
\item Carry out simulations and extract correlation functions.
\item Analyze results in an attempt to identify states. 
\end{itemize}

Once reasonable numbers are obtained, with the
accumulated experience one can return to the most interesting problem, that is to construct an explicit
example of a RG transformation restoring translational
invariance at its fixed point and clarify what type
of tuning is necessary to induce the flow in the IR to
the Wilson-Fisher fixed point. 

Even if a good analogue of a Wilsonian RG is not found, the desired continuum limit is still likely to emerge on the basis of results obtained so far. In lieu of a good RG analogue, the classification of corrections subleading in $N$, the number of vertices on one spherical shell, needs to be
addressed directly. The natural guess is that the asymptotics as $N\to\infty$ can be expressed  by an effective continuum action whose leading term alone produces the continuum radially quantized target CFT and corrections are ordered by dimensions of scalar primaries. This can hold only up to some non-zero fraction of $N$, $\zeta N$. The states with dimensions lower than $\zeta N$ fall into angular momentum multiplets which stay unsplit under $I_h$. The assumption is that $\zeta$ stays larger than zero as $N\to\infty$. 

At this point it is premature to present generalizations 
of lattice radial quantization to non-scalar fields~\cite{future1}.

\begin{acknowledgments}
This research was supported in part by the
NSF under award PHY-1415525. Figures have been produced with {\it Mathematica}.
\end{acknowledgments}


\begin{thebibliography}{5}
\bibitem{ourplb} R.~C.~Brower, G.~F.~Fleming, H.~Neuberger, Phys. Lett. B 721 (2013) 299.
\bibitem{cardy} J.~L.~Cardy, J. Phys. A 18 (1985) L757.
\bibitem{randlat} K.~I.~Macrae, Phys. Rev. D. 23 (1981) 886; N.~H. Christ, R.~Friedberg, T.~D.~Lee, Nucl. Phys. B202 (1982) 89. 
\bibitem{fix} G. Strang, G. J. Fix, ``An Analysis of the Finite Element Method'', 
Prentice-Hal, 1973. 
\bibitem{brower2013} 
  R.~C.~Brower, M.~Cheng and G.~T.~Fleming,
  PoS LATTICE {\bf 2013}, 335. 
\bibitem{brower2014} 
  R.~C.~Brower, Talk delivered at the 32nd International Symposium on Lattice Field Theory, 23-28 June, 2014, Columbia University New York, NY.
  \bibitem{cheeger} J. Cheeger, W. M{\" u}ller, R. Schrader, Comm. Math Phys. 92 (1984) 405.
  \bibitem{wardetzky} K. Hildebrandt, K. Pothier,
  M. Wardetzky, Geom. Dedicata 123 (2006) 89.
  \bibitem{sorkin} R. Sorkin, J. of Math. Phys. 
  (1975) 2432.  
  \bibitem{shallow} G.~R.~Stuhne, W.~R.~Peltier, 
  J. of Comp. Phys. 1948 (1999) 23. 
  \bibitem{imaging} M.~K.~Chung, K.~J.~Worsley, S.~Robbins, A.~C.~Evans, in IEEE, ``Conference on Computer
  Vision and Pattern Recognition (CVPR)'', vol I (2003) 467. 
  \bibitem{binder} K.~Binder, Z.~Phys. B 43 (1981) 119.
    \bibitem{vicari} A.~Pelissetto, E.~Vicari, Phys. Rept. 368 (2002) 549.
   \bibitem{weigel} M.~Weigel, W.~Janke, Phys. A 281 (2000) 287.
   \bibitem{cones} B.~S.~Kay, U.~M.~Studer, Comm. Math. Phys. 139 (1991) 103.
   \bibitem{cosmic} V.~Frolov, D.~Fursaev, Class., Qu.,
   Grav. 18 (2001) 1535. 
     \bibitem{cub}  K.~Atkinson, W.~Han, 
  ``Spherical Harmonics and Approximations on the
  Unit Sphere: An Introduction'', Springer 2012.
  \bibitem{smear} R.~Narayanan, H.~Neuberger,
  JHEP {\bf 0603}, 064 (2006).
  \bibitem{gaussquad} C.~Ahrens, G.~Beylkin, Proc. R. Soc.
  A 465 (2009) 3103; N.~L.~Fern\'{a}ndez, Elec. Trans. on Num. Anal. 19 (2005) 84; J.~Prestin, D.~Ro{\c s}ca, J. of Approx. Th. 142 (2006) 1.
  \bibitem{gallavotti} G. Benfatto, G. Gallavotti, 
  ``Renormalization Group'', Princeton University Press (1995). 
  \bibitem{belkin} M.~Belkin, P.~Niyogi, Neural Computation 15 (2003) 1373.
  \bibitem{dziuk} G. Dziuk, ``Finite elements for the Beltrami operator on
  arbitrary surfaces''. In ``Partial Differential Equations and Calculus of Variations'', S. Hildebrandt and R. Leis, eds., Vol 13, Lecture Notes in Mathematics, Springer (1988) 142; G. Dziuk, C. M. Elliott, Acta Numerica (2013) 289.
  \bibitem{overlap} H. Neuberger, Phys. Lett. B417 (1998) 
  141.
  \bibitem{mina} S.~Minakshisundaram, $\overset{\circ}{\rm A}$. Pleijel, 
  Can. J. Math. I (1949) 242.
  \bibitem{jaffe} A.~Jaffe and G.~Ritter,
    Commun.\ Math.\ Phys.\  {\bf 270} (2007) 545; arXiv:0704.0052 [hep-th].
    \bibitem{hsu} E.~S.~Hsu, ``Stochastic Analysis on
    Manifolds'', AMS (2001).
    \bibitem{dressel} M. S. Dresselhaus, G. Dresselhaus, A. Jorio,
    ``Group Theory, Application to the Physics of Condensed 
    Matter'', Springer (2008).
    \bibitem{gard} N. B. Backhouse, P. Gard, J. Phys. A: Math. Nucl. Gen. 7 (1974) 2101.
    \bibitem{sobolev} S.~L.~Sobolev, Dokl. Akad. Nauk SSSR 146 (1962) 41,
    Dokl. Akad. Nauk SSSR 146 (1962) 310,
    Dokl. Akad. Nauk SSSR 146 (1962) 770; J.~M.~Goethals, J.~J.~Seidel, 
    ``Cubature Formulae, Polytopes, and Spherical Designs'', in ``The
    Geometric Vein'', C. Davis et. al. (eds), Springer-Verlag, NY, 1981.  
    \bibitem{flato} L.~Flatto, Bull. Amer. Math. Soc., 74 (1968) 730.
    \bibitem{coxeter}H.~S.~M. Coxeter, ``Regular Polytopes'', Dover (1973).
    \bibitem{humphreys} J.~E.~Humphreys, ``Reflection Groups and Coxeter Groups'', Cambridge University Press (1990). 
    \bibitem{klein} F.~Klein, ``Vorlesungen \"{u}ber das Ikosaeder und die Aufl\"{o}sung der Gleichungen f\"{u}nften Grades'', Leipzig (1884). 
    \bibitem{f4} H.~Neuberger,
  Phys.\ Lett.\ B {\bf 199} (1987) 536;
  U.~M.~Heller, M.~Klomfass, H.~Neuberger and P.~M.~Vranas,
  Nucl.\ Phys.\ B {\bf 405} (1993) 555.
  \bibitem{cromwell} P.~R.~Cromwell, ``Polyhedra'', Cambridge University Press (1997).
    \bibitem{future1} H. Neuberger, in preparation.
    \bibitem{jackiw} S.~Fubini, A.~J.~Hanson, R.~Jackiw, Phys. Rev. D 7 (1973) 1732.
    \bibitem{ruhl} K.~Lang, W.~R\"{u}hl, Nucl. Phys. B402 (1993) 573.
    \bibitem{zinn} J.~Zinn-Justin, ``Quantum Field Theory and Critical Phenomena'', Oxford University Press (1999).

\end{thebibliography}
\end{document}